\documentclass[12pt,a4paper]{article}
\usepackage[utf8]{inputenc}
\usepackage{amsmath}
\usepackage{amsfonts}
\usepackage{amssymb}
\usepackage{bm}
\usepackage{pdflscape}
\usepackage[margin=1in,footskip=0.25in]{geometry}
\usepackage{graphicx}
\usepackage{multirow}
\usepackage[table,xcdraw]{xcolor}

\begin{document}
\noindent \textbf{\Large{$\mathcal{R}=P(Y<X)$ for unit-Lindley distribution: inference with an application in public health}}\\
\\
Aniket Biswas\textsuperscript{a} \,\,and\,\, Subrata Chakraborty\textsuperscript{b}\\
Department of Statistics\\
Dibrugarh University\\
Dibrugarh-786004\\
India\\
\\
Email:\,\,\textsuperscript{a}\textit{biswasaniket44@gmail.com} \,\,\,\,\,\textsuperscript{b}\textit{subrata\_stats@dibru.ac.in}

\section*{Abstract}
The unit-Lindley distribution was recently introduced in the literature as a viable alternative to the Beta and the Kumaraswamy distributions with support in $(0,1)$.  This distribution enjoys many virtuous properties over the named distributions. In this article, we address the issue of parameter estimation from a Bayesian perspective and study relative performance of different estimators through extensive simulation studies. Significant emphasis is given to the estimation of stress-strength reliability employing classical as well as Bayesian approach. A non-trivial useful application in the public health domain is presented proposing a simple metric of discrepancy.    
\section*{keywords}
unit-Lindley distribution, stress-strength reliability, conjugate prior, Metropolis-hastings algorithm, Weighted gamma distribution, Kummer's U-function.   
\section{Introduction}
The importance and relevance of probability models in the support $(0,1)$ is well established in applied statistics to model certain characteristics such as, scores of some ability tests, different indices and rates, which lie in the interval $(0, 1)$.
The Unit Lindley distribution was introduced by Mazucheli et al.(2019) as  a new distribution in the unit interval $ (0,1) $  with many interesting features. Its probability density function (p.d.f) and cumulative distribution function (c.d.f) is given respectively by:
\begin{equation} 
f(x\mid \theta )=\dfrac{\theta ^{2}}{1+\theta }\,(1-x)^{-3}\,\exp \left( -%
\dfrac{\theta \,x}{1-x}\right) ,\qquad 0<x<1,\,\theta >0.  \label{eq:pdf}
\end{equation}

\begin{equation}
F(x\mid \theta )=1-\left( 1-\dfrac{\theta \,x}{(1+\theta )\,(x-1)}\right)
\,\exp \left( {-\dfrac{\theta \,x}{1-x}}\right) ,\qquad 0<x<1,\,\theta >0.
\label{eq:cdf}
\end{equation}
This is a one parameter unimodal distribution with many advantageous characteristics like closed form expressions
for cumulative distribution function (c.d.f), quantile function and simple expression for lower order moments that the other distributions defined in the interval $(0,1)$ do not posses. Mazucheli et al. (2018) justified the utility of this distribution as a viable competitor of the existing distributions especially the beta and Kumaraswamy distributions by providing theoretical results and practical application with illustrations.\\

Stress-strength reliability is an important probabilistic measure used regularly in the field of  quality engineering, medical statistics, econometrics,  and other branches of applied statistics. If $X$ stands for the strength of a system or component which is subjected to a stress $Y$, then the stress-strength reliability $\mathcal{R}$ is defined as the probability that $Y$ is less than $X$, that is, stress is less than the strength. In our case, both $X$ and $Y$ are proper fractions, like per unit rate of any event of interest, proportion of some categories, relative grades of any inspection or test etc. In all such situations, it may be of interest to monitor the probability that one is smaller or larger than the other given by the following expression when $X$ and $Y$ are independent.
\begin{equation*}
\label{R}
R=P\left( Y<X\right) =\int_{0}^{1}\,f_{X}(x\mid \theta _{1})\,F_{Y}(x\mid
\theta _{2})\,dx  \notag \\
\end{equation*}
Stress-strength reliability estimation has attracted many authors (For details see Krishna et al.(2017) and references therein). However, this problem for distributions with support $(0,1)$ is rare in the literature. The works we came across is for Kumaraswamy distribution, discussed in detail by Nadar et al.(2014) and for Topp-Leone distribution in Genc(2013), where the author applied $\mathcal{R}$ to evaluate efficiency of two different algorithms based on unit capacity factor. Many survey data available in public health domain are often reported in percentages or equivalently in proportions. Unit-Lindley has been shown to model proportion data quite well in section 5 of Mazucheli et al.(2019). Here, using $\mathcal{R}$, we develop a metric to quantify discrepancy between two geographical regions with respect to certain public health indicator as follows:\\

Consider two non-overlapping regions: $A$ comprising $n$ distinct sub-regions and $B$ with $m$ sub-regions , say. Suppose, proportion of population with access to a certain indicator ($I$) for these sub-regions are available. The $m$ observations available for $A$ are iid copies of $Y$ and those for $B$ are copies of $X$. Denoting $\mathcal{R}(A,B)=P(Y<X)$, a discrepancy measure between $A$ and $B$ with respect to $I$ is defined as,
\begin{equation}
\mathcal{D}(A,B)=\mathcal{R}(B,A)-\mathcal{R}(A,B)\nonumber\\
\end{equation}    
Unlike in reliability engineering, $\mathcal{R}$ has not been used in public health at all for lack of intuitive appeal. On the contrary, $\mathcal{D}(A,B)$ with the following desirable properties can easily manifest the prevailing situation:\\
\textit{
\begin{itemize}
    \item \textbf{1:} $-1\leq\mathcal{D}(A,B)\leq 1$ for any $A$, $B$.
    \item \textbf{2:} For any $A$, $\mathcal{D}(A,A)=0$.
    \item \textbf{3:} For any $A$ and $B$, $\mathcal{D}(A,B)=-\mathcal{D}(B,A)$.
    \item \textbf{4:} For any $A$, $B$ and $C$, $\mathcal{D}(A,B)>(<)0$, $\mathcal{D}(B,C)>(<)0$ $\implies$ $\mathcal{D}(A,C)>(<)0$.  
    \end{itemize}}  
$\mathcal{D}(A,B)=0$ will indicate almost similar situation in the two regions while a value of the same indicates better situation in $A$ than in $B$. $\mathcal{D}(A,B)=+1/-1$ are two obvious extreme conditions emphasizing prevalence of a strong discrepancy.\\
 
Rest of this article is organized as follows: In Section \ref{Estimation Theta}, we discuss the problem of estimating $\theta$. Exact expressions for the Bayes estimator under both conjugate as well as flat prior are obtained in terms of  Kummer’s confluent Hypergeometric function. In Section \ref{Estimation R}, main results for maximum likelihood estimation, uniformly minimum variance unbiased estimator (UMVUE) and Bayes estimation for stress-strength reliability $R=P(Y<X)$ under independent set-up are developed. Both these sections provide computational algorithms. The next section is devoted to simulation set up, results and interpretations to asses the performance of the proposed estimators. In section 5, an application of the proposed metric is illustrated through basic and safely managed drinking water services data made available by WHO (http://apps.who.int/gho/data/view.main.WSHWATERv?lang=en). We conclude with some discussion and directions for future extension in section 6. An appendix of relevant tables is provided at the end.\\

Through out the rest of the text we write $L(\theta)$ and $UL(\theta)$ to denote Lindley and unit-Lindley distribution with the parameter $\theta$.
\section{Estimation of $\theta$}\label{Estimation Theta}
In this section, we discuss maximum likelihood estimation of $\theta$ (Mazucheli et al., 2019) and introduce estimation of $\theta$ under Bayesian paradigm. We find the conjugate-prior family for $\theta$ and derive closed form expression of posterior as a weighted Gamma distribution. A computational algorithm for sampling from posterior distribution is outlined. 
\subsection{Classical approach}
\subsubsection{MLE}
Let $\bm{X}=(X_{1},X_2,\ldots ,X_{m})$ be a random sample of size $m$ from $UL(\theta)$ with p.d.f. \eqref{eq:pdf}. Then, for realized  $\bm{X}:=\bm{x}%
=(x_{1},\ldots ,x_{n})$, the log-likelihood function of $\theta $ can be
written as:
\begin{equation}
\mathcal{L} (\theta \mid \bm{x})=\prod_{i=1}^{m}(1-x_i)^{-3}\left(\frac{\theta^2}{1+\theta}\right)^m \exp(-\theta \,t(\bm{x})) \label{eq:like}
\end{equation}%
where $t(\bm{x})=\sum\limits_{i=1}^{m}\,\dfrac{x_{i}}{1-x_{i}}$.
which gives:
\begin{equation}
\widehat{\theta }_{ML}=\frac{1}{2\,t(\bm{x})}\left[ m-t(\bm{x})+\sqrt{t(\bm{x}%
	)^{2}+6\,m\,t(\bm{x})+m^{2}}\right] .  \label{eq:mle}
\end{equation}
Mazucheli et al.(2018) provides the following asymptotic distribution of $\widehat{\theta}_{ML}$:
$$
\sqrt{m} \,(\widehat{\theta}_{ML}- \theta) \sim AN\left(0, \sigma^2(\theta) \right)
$$
where, \, $\sigma^2(\theta)=\dfrac{\theta^{2}\,(1+\theta )^{2}}{(\theta ^{2}+4\,\theta +2)}$ and thus
constructed asymptotic confidence interval for $\theta$. Table 1 of the same paper reveals that, performance of maximum likelihood estimator of $\theta$ is satisfactory and equivalent to method of moments estimator while for small and moderate sample sizes bias corrected estimator performs better. However, due to its desirable properties such as invariance, consistency, asymptotic normality etc., $\widehat{\theta}_{ML}$ should still be preferred for further inferential work on complex parametric functions.
\subsubsection{UMVUE}
For unit-Lindley distribution, $t(\bm{x})$ is a complete sufficient statistic and is distributed as $W=\sum_{i=1}^m W_i$, where $W_i\sim L(\theta)$ for $i=1,2,...,m$ and are independent. From Zakerzadeh and Dolati(2009), the p.d.f. of $W$ is seen as follows:
\begin{equation}
f_{W}(w)=\left(\frac{\theta^2}{1+\theta}\right)^m\,\sum_{k=0}^m {m\choose k}\frac{w^{2m-k-1}\exp{(-\theta\,w)}}{\Gamma{(2m-k)}}\quad\textrm{for}\quad w>0\label{eq:pdfsumlind}
\end{equation}
Since unbiased estimator of $\theta$ is not known, direct construction of UMVUE using Lehmann-Scheffe's theorem is not available. Mazucheli et al.(2018) indicates construction of UMVUE by bias correction of the estimator in \eqref{eq:mle}. Maiti and Mukherjee(2018) used empirical cdf to construct UMVUE of cdf and pdf of Lindley distribution. Similar routine can be followed for unit-Lindley distribution. 

\subsection{Bayesian approach}
In Mazucheli et al.(2019), the authors mentioned about Bayesian estimation of the parameter $\theta$, but to the best of our knowledge no such work has so far been reported. As mentioned above in this paper,  we consider the problem of Bayesian estimation of $\theta$ with improper and conjugate priors. Prior sensitivity is also assesed through extensive simulation studies for motivating practitioners to apply the proposed method safely. \\

Ignoring the data-dependent part in \eqref{eq:like}, we rewrite the likelihood function as follows:  
\begin{equation}
\mathcal{L} (\theta \mid \bm{x})\propto\left(\frac{\theta^2}{1+\theta}\right)^m \exp(-\theta \,t(\bm{x})) \label{eq:proplike}
\end{equation}%

\subsubsection{Conjugate Prior}
In section 2.3 of Mazucheli et al.(2019) the concerned distribution is shown to be a member of one-parameter exponential family which facilitates one to construct family of conjugate priors (see Robert, 2007). Accordingly the constructed family of conjugate priors for the unit-Lindley distribution is the following:
\begin{equation}
    \mathbf{\pi}(\theta)\propto \frac{\theta^{p-1}\exp{(-\theta\,\alpha)}}{(1+\theta)^{\beta}}\quad\textrm{for}\quad \theta>0\label{eq:prior}
\end{equation}
where, $\alpha,p>0$ and $\beta\geq 0$.\\

The above density can  easily be identified as a weighted gamma distribution for which the normalizing constant can be seen as the confluent hypergeometric function of the second kind (Abramowitz and stegun, 1972) popularly known as Kummer's function. It has a convenient Maclaurine's series expansion. Henceforth density in \eqref{eq:prior} will be denoted by $WG(\alpha,\beta,p)$. Taking product of the likelihood function in \eqref{eq:proplike} with the prior in \eqref{eq:prior}, the posterior is found as follows:    

\begin{equation}
    \mathbf{\pi}(\theta|\bm{x})\propto \frac{\theta^{2m+p-1}\exp{[-\theta\,\{\alpha+t(\bm{x})\}]}}{(1+\theta)^{m+\beta}}\quad\textrm{for}\quad \theta>0\label{eq:posterior}
\end{equation}
As obvious the the density in \eqref{eq:posterior} is also a member of weighted-gamma family: $WG(\alpha+t(\bm{x}),m+\beta,2m+p)$. The marginal density of data vector $\bm{X}$ can be obtained by integrating out the posterior density w.r.t $\theta$ which takes the following form:
\begin{equation}
m(\bm{x})=\frac{\Gamma{(2m+p)}}{\Gamma{p}}\frac{U(2m+p,m+p+1-\beta,\alpha+t(\bm{x}))}{U(p,p+1-\beta,\alpha)}\label{eq:marginal}
\end{equation}
where, $\Gamma$ denotes the gamma function and $U$ denotes Kummer's U function (see,Abramowitz and stegun, 1972). Considering squared error loss function makes the posterior mean to be the Bayes' estimator which in this case turns out to be:
\begin{equation}
E(\theta|\bm{x})=\frac{\Gamma{(2m+p+1)}}{\Gamma{(2m+p)}}\,\frac{U(2+p,p+2-\beta,\alpha+t(\bm{x}))}{U(2m+p,m+p+1-\beta,\alpha+t(\bm{x}))}\label{eq:postmeanconj}
\end{equation}
From the literature review it is apparent that, there is a natural tendency to use gamma prior when parameter of interest has positive support (see Krishna et al., 2017 and Nadar et al., 2014 among others). In fact, gamma prior is a particular case of the proposed frame-work when $\beta=0$ and we denote this by $G(\alpha,p)$. Even in this situation, posterior of $\theta$ is still in weighted gamma family implying that, gamma distribution plays the role of semi-conjugate prior: $WG(\alpha+t(\bm{x}),m,2m+p)$. 
\\

As the prior is a modification of gamma distribution with simple weights, subjective belief regarding the parameter of interest can be meaningfully captured. Use of conjugate prior, provide computational tractability and enhance transparency in updation of prior through likelihood.\\

In full Bayesian approach, the hyper-parameters are purely in the hands of the practitioner. As a trade-off between classical and Bayesian paradigm, in empirical Bayes' approach, the choice of hyper-parameters are data-driven. In this approach, the maximum likelihood estimates of the hyper-parameters from the marginal distribution of the data are plugged into the expression of posterior mean to obtain Bayes' estimator. For more details, see Efron(2012). In our case, the marginal distribution given in \eqref{eq:marginal} is quite complex and direct computation of maximum likelihood estimates is ruled out while numerical optimization of $m(\bm{x})$ w.r.t $\alpha,\,\beta,\,p$ provides no stable solution. Even our attempt to derive the estimates of hyper-parameters by other methods proved futile. Due to the above issues, we do not pursue this method further in this work.  
  
\subsubsection{Flat Prior}
In situations where no information regarding the parameter of interest is available, one may use flat priors to express indifference. Obviously, a number of conventional choices of flat priors are in use of which the present work uses the basic one:
\begin{equation}
\mathbf{\pi}(\theta)\propto 1 \quad\textrm{for}\quad \theta>0 \label{eq:imp_prior}    
\end{equation}
The corresponding posterior is $WG(t(\bm{x}),m,2m+1)$ which indicates that the assumed improper prior is also a semi-conjugate one. It should be noted that, even with improper prior, the corresponding posterior is proper and hence under squared error loss function, the Bayes' estimator is given by:
\begin{equation}
E(\theta|\bm{x})=(2m+1)\frac{U(2m+2,m+3,t(\bm{x}))}{U(2m+1,m+2,t(\bm{x}))}\label{eq:postmeanimp}    
\end{equation}

\subsubsection{Computational Methods}
Posterior mean of parameter is the optimum estimator under squared error loss function. For conjugate and improper cases, mean of posterior distribution is given in \eqref{eq:postmeanconj} and \eqref{eq:postmeanimp}, respectively. We denote Bayes' estimator of $\theta$ for improper prior by $\widehat{\theta}_I$ whereas $\widehat{\theta}_A$ denotes the same for conjugate prior. Prior-sensitivity is investigated through simulation studies by miss specifying the hyper-parameters $\alpha,\beta,p$ as $\alpha^\prime,\beta^\prime,p^\prime$. The corresponding estimator is denoted by $\widehat{\theta}_M$. \\

To assess the posterior density of $\theta$, one may need to draw sample. For the priors considered in this work posterior family remains the same, weighted gamma distribution. Metropolis-Hastings (MH) algorithm is widely accepted and used for drawing sample from un-normalized posterior using MCMC. We briefly state the algorithm below for single parameter:
\textit{
\begin{itemize}
    \item \textbf{Step 1:} Initialize $\theta=\theta^{(0)}$.
    \item \textbf{Step 2:} Draw a candidate for $\theta^{*}\sim h(\theta|\theta^{(0)},\bm{x})$, $h$ being the proposal density.
    \item \textbf{Step 3:} Compute the acceptance ratio using target density $\pi$ and proposal $h$ as:
    \begin{equation*}
        r=\frac{\pi(\theta^{*}|\bm{x})}{\pi(\theta^{(0)}|\bm{x})}\,\frac{h(\theta^{(0)}|\theta^{*})}{h(\theta^{*}|\theta^{(0)})}
    \end{equation*}
    \item \textbf{Step 4:} Draw $u\sim Uniform(0,1)$.
    \item \textbf{Step 5:} Set,
    \begin{equation*}
      \theta^{(1)}=\begin{cases} 
      \theta^{*}\,\,\,\,\, \textrm{if}& u<r\\
      \theta^{(0)}\,\,\,\,\, \textrm{if}& u>r\\
   \end{cases}
    \end{equation*}
    \item \textbf{Step 6:} $\theta^{(0)}\longleftarrow\theta^{(1)}$.
    \item \textbf{Step 7:} Repeat \textbf{Step 2} to \textbf{Step 6} $c$ times with burn in $b$ for $b\leq c$ to obtain $c-b+1$ sample observations.
    \end{itemize}}
Several improvements of the basic MH-algorithm have been proposed and got implemented in R-software. The present work uses \textit{MHadaptive} package from \textbf{CRAN} repository (see Chivers,2015) with non-default burn in length $b=5000$ and chain-length $c=14999$. 
\section {Estimation of $\mathcal{R}$}\label{Estimation R}
Let us consider independent random samples, $\bm{X}=(X_1,X_2,...,X_m)$  of size $m$ from UL($\theta_1$) and $\bm{Y}=(Y_1,Y_2,...,Y_n)$ of size $n$ from UL($\theta_2$) with $m>n$ such that,
\begin{equation}
\frac{n}{m}=q\in(0,1) \qquad\textrm{for}\qquad m,n\longrightarrow \infty \label{eq:mnlimit}
\end{equation}
Under this set-up, the stress strength reliability parameter can be expressed as,
\begin{eqnarray}
\mathcal{R}&=&\dfrac{\theta _{2}^{2}\,\left(\theta_1\, \theta _{2}^{2}+2\,\theta
	_{1}^2\,\theta _{2}+\theta _{1}^{3}+\theta _{2}^{2}+4\,\theta _{1}\,\theta
	_{2}+3\,\theta _{1}^{2}+\theta _{2}+3\,\theta _{1}\right) }{(\theta
	_{1}+\theta _{2})^{3}\,( 1 + \theta _{2})\,(1+\theta _{1})}\nonumber\\
	&=& g(\theta_1,\theta_2)\,\, ,\textrm{say}\label{eq:R}
\end{eqnarray}
The polynomial in the denominator of \eqref{eq:R} is non-zero and $g(\theta_1,\theta_2)$ being the ratio of two polynomials, is a continuous function over $(0,1)$. In what follows, we will consider some classical and Bayesian methods for estimating $\mathcal{R}$.
\subsection {Classical approach}
\subsubsection{MLE}
In view of the functional invariance property, the MLE of $\mathcal{R}$ is given by,
\begin{eqnarray}
\label{RML}
\widehat{\mathcal{R}}_{ML}&=&\dfrac{\widehat{\theta _{2}}^{2}\,\left(\widehat{\theta _{1}}\, \widehat{\theta _{2}}^{2}+2\,\widehat{\theta _{1}}^2\,\widehat{\theta _{2}}+\widehat{\theta _{1}}^{3}+\widehat{\theta _{2}}^{2}+4\,\widehat{\theta _{1}}\,\widehat{\theta _{2}}+3\,\widehat{\theta _{1}}^{2}+\widehat{\theta _{2}}+3\,\widehat{\theta _{1}}\right) }{(\widehat{\theta _{1}}+\widehat{\theta _{2}})^{3}\,( 1 + \widehat{\theta _{2}})\,(1+\widehat{\theta _{1}})} \nonumber\\
&=& g(\widehat{\theta}_1,\widehat{\theta}_2)\label{eq:Rhat}
\end{eqnarray}
where, $\widehat{\theta}_1$ and $\widehat{\theta}_2$ are respective MLE's of $\theta_1$ and $\theta_2$.  
From section (2.1.1) we have,
\begin{eqnarray*}
\sqrt{m}(\widehat{\theta}_1-\theta_1)\sim AN\left(0,\sigma^2(\theta_1)\right)\\
\sqrt{n}(\widehat{\theta}_2-\theta_2)\sim AN\left(0,\sigma^2(\theta_2)\right)
\end{eqnarray*}
Let us denote,
\begin{equation*}
    \bm{\theta}=(\theta_1,\theta_2)\qquad\textrm{and}\qquad\widehat{\bm{\theta}}=(\widehat{\theta}_1,\widehat{\theta}_2)
\end{equation*}
Clearly,
\begin{equation*}
 \widehat{\bm{\theta}}\sim AN_2\left(\bm{\theta},\Sigma(\bm{\theta})\right)   
\end{equation*}
where,
$$\Sigma(\bm{\theta})= \begin{bmatrix}
\sigma^2(\theta_1) & 0 \\
0 & \sigma^2(\theta_2) 
\end{bmatrix}  $$
As mentioned earlier, $g$ is continuous and applying $\delta$-method we get,
\begin{equation}
    \sqrt{m}\,\left(g\left(\widehat{\theta}_1,\widehat{\theta}_2\right)-g(\theta_1,\theta_2)\right)\sim AN\left(0,\sigma_*^2(\theta_1,\theta_2)\right)\label{eq:rml}
\end{equation}
Here, $\sigma_*^2(\theta_1,\theta_2)$ can be obtained using the following:
\begin{eqnarray}
\sigma_*^2(\theta_1,\theta_2)=  \begin{bmatrix}
\frac{\delta g(\theta_1,\theta_2)}{\delta \theta_1} & \frac{\delta g(\theta_1,\theta_2)}{\delta \theta_2}\end{bmatrix} \,\, \Sigma(\bm{\theta})\,\,\begin{bmatrix}
\frac{\delta g(\theta_1,\theta_2)}{\delta \theta_1} & \frac{\delta g(\theta_1,\theta_2)}{\delta \theta_2}\end{bmatrix}^\prime \label{eq:delmethod}
\end{eqnarray}
We introduce the following notations to state the final expression which is quite messy in our case.
\begin{eqnarray*}
a_1&=&\frac{\theta_2^2\left(3+6\theta_1+3\theta_1^2+4\theta_2+4\theta_1\theta_2+\theta_2^2\right)}{\left(1+\theta_1\right)\left(1+\theta_2\right)\left(\theta_1+\theta_2\right)^3}\\
a_2&=&\frac{3\theta_2^2\left(3\theta_1+3\theta_1^2+\theta_1^3+\theta_2+4\theta_1\theta_2+2\theta_1^2\theta_2+\theta_2^2+\theta_1\theta_2^2\right)}{\left(1+\theta_1\right)\left(1+\theta_2\right)\left(\theta_1+\theta_2\right)^4}\\
a_3&=&\frac{\theta_2^2\left(3\theta_1+3\theta_1^2+\theta_1^3+\theta_2+4\theta_1\theta_2+2\theta_1^2\theta_2+\theta_2^2+\theta_1\theta_2^2\right)}{\left(1+\theta_1\right)^2\left(1+\theta_2\right)\left(\theta_1+\theta_2\right)^3}\\
e&=&\left(a_1-a_2-a_3\right)^2\\
b_1&=&\theta_1^2\left(1+\theta_1\right)^2\\
b_2&=&\theta_2^2\left(1+\theta_2\right)^2\\
c_1&=& 2+4\theta_1+\theta_1^2\\
c_2&=&q\left(2+4\theta_2+\theta_2^2\right)\\
d_1&=&\frac{b_1\,e}{c_1}\\
d_2&=&\frac{b_2\,e}{c_2}
\end{eqnarray*}
With these notations, we get from \eqref{eq:delmethod}
\begin{equation}
    \sigma_{*}^2(\theta_1,\theta_2)=m\,(d_1+d_2)\label{eq:sigstarsq}
\end{equation}
For all practical purposes, one can substitute the corresponding MLE's in place of $\theta_1$ and $\theta_2$ to obtain
\begin{equation*}
    \widehat{\sigma}_*^2(\theta_1,\theta_2)=\sigma_*^2(\widehat{\theta}_1,\widehat{\theta}_2)
\end{equation*}
Thus, using asymptotic normality of $\widehat{\mathcal{R}}_{ML}$ given in \eqref{eq:rml} $100\,(1-\alpha)\%$ confidence interval can be constructed as follows:
\begin{equation*}
    CI(\mathcal{R})=\left[\widehat{\mathcal{R}}_{ML} -\tau_{\alpha/2}\,\frac{\widehat{\sigma}_*^2(\theta_1,\theta_2)}{\sqrt{m}}\,\,,\,\,\widehat{\mathcal{R}}_{ML}+\tau_{\alpha/2}\,\frac{\widehat{\sigma}_*^2(\theta_1,\theta_2)}{\sqrt{m}}\right]
\end{equation*}
where, $\tau_\alpha$ denotes the upper-$\alpha$ point of standard normal distribution.
\subsubsection{UMVUE}
As in subsection 2.1.2, 
\begin{eqnarray*}
t(\bm{x})&=&W=\sum_{i=1}^m W_i\qquad\textrm{where}\qquad W_i\sim L(\theta_1)\quad\textrm{for}\quad i=1,2,...,m.\\
t(\bm{y})&=&V=\sum_{i=1}^n V_i\qquad\,\,\,\textrm{where}\,\,\,\qquad V_i\sim L(\theta_2)\quad\textrm{for}\quad i=1,2,...,n.
\end{eqnarray*}
$(W,V)$ is jointly complete sufficient for $(\theta_1,\theta_2)$ and unlike for $\theta$ in case of UL($\theta$) here, an unbiased estimator for $\mathcal{R}$ can easily be constructed as
\begin{equation*}
    \psi(X_1,Y_1)=\begin{cases}
    1\,\,\,if\,&Y_1<X_1\\
    0\,\,\,\,\,\, &otherwise
    \end{cases}
\end{equation*}
It is important to note that, $f(x)=x/(1-x)$ is an increasing function in $x\in (0,1)$. Thus, the above indicator can be restated as 
\begin{equation*}
    \phi(W_1,V_1)=\begin{cases}
    1\,\,\,if\,&V_1<W_1\\
    0\,\,\,\,\,\, &otherwise
    \end{cases}
\end{equation*}
 One can imitate the steps in Al-Mutairi et al.(2013) to easily find the expression for UMVUE of $\mathcal{R}$. Table 1 of Al-Mutairi et al.(2013) clearly indicates that, MLE beats UMVUE in terms of MSE in case of Lindley. Moreover, the complex nature of UMVUE makes computaion expensive. It is reasonable to expect a similar situation in case of unit-Lindley, we refrain from computational aspect of UMVUE.  
\subsection{Bayesian approach}
Prior belief on $\theta_1$ and $\theta_2$ suffices need for prior on $\mathcal{R}$. Thus, priors mentioned through subsection 2.2 are kept intact for both $\theta_1$ and $\theta_2$ independently, making way for Bayes' estimation of $\mathcal{R}$. We extend the conjugate set-up for single parameter to the case of two parameters as follows:
\begin{eqnarray*}
 \theta_1\sim WG(\alpha_1,\beta_1,p_1)\\
 \theta_2\sim WG(\alpha_2,\beta_2,p_2)
\end{eqnarray*}
Similarly, the improper set-up can be extended for two-sample situation
\begin{equation*}
    \pi(\theta_1)\propto 1\qquad\textrm{and}\qquad \pi(\theta_2)\propto 1
\end{equation*}
Whatever be the set-up, as previously mentioned, posteriors of both the parameters are in weighted gamma family. Posterior of the parameters under conjugate set-up:
\begin{eqnarray*}
\theta_1|\bm{x}&\sim&WG(\alpha_1+t(\bm{x}),m+\beta_1,2m+p_1)\\
 \theta_2|\bm{y}&\sim& WG(\alpha_2+t(\bm{y}),n+\beta_2,2n+p_2)
\end{eqnarray*}
and for improper prior,
\begin{eqnarray*}
 \theta_1|\bm{x}&\sim& WG(t(\bm{x}),m,2m+1)\\
 \theta_2|\bm{y}&\sim& WG(t(\bm{y}),n,2n+1)
\end{eqnarray*}
As we consider both the samples and priors on the parameters to be independent, posteriors of $\theta_1$ and $\theta_2$ are obviously independent of each other. Given the expression in \eqref{eq:R}, it is near impossible to deduct the posterior distribution of $\mathcal{R}$ and thus the posterior mean. So a good strategy would be to draw observations from the posterior distribution of $\mathcal{R}$ and computing the posterior mean based on a large sample. Use of efficient algorithm warrants closeness of the true mean with the simulation based mean, which is quite common in Bayesian approach.  We present algorithm for the same below:
\textit{
\begin{itemize}
    \item \textbf{Step 1:} Make an array of size $k$ for posterior sample of $\theta_1$ and $\theta_2$ using algorithm given in subsection 2.2.3.  
    \item \textbf{Step 2:} For all $i=1,2,...,k$ compute $R^{(i)}=g(\theta_1^{(i)},\theta_2^{(i)})$ and store in array $\bm{R}$.
    \item \textbf{Step 3:} Calculate mean of $\bm{R}$.
\end{itemize}}
\noindent Here, $k=10000$ which follows from the choice of $b$ and $c$ given in subsection 2.2.3 which performs satisfactory in simulation studies. As mentioned in subsection 2.2.3 for $\theta$, We denote Bayes' estimator of $\mathcal{R}$ for improper prior by $\widehat{\mathcal{R}}_I$ whereas $\widehat{\mathcal{R}}_A$ denotes the same for conjugate prior. Prior-sensitivity is investigated through simulation studies by miss specifying the hyper-parameters $\alpha_1,\beta_1,p_1$ as $\alpha_1^\prime,\beta_1^\prime,p_1^\prime$ and $\alpha_2,\beta_2,p_2$ as $\alpha_2^\prime,\beta_2^\prime,p_2^\prime$. The corresponding estimator is denoted by $\widehat{\mathcal{R}}_M$.
\section{Simulation study}\label{simulation}
Despite the discussion regarding performance of MLE for $\theta$ in Mazucheli et al.(2018), we extend the experiment for comparison of the same with Bayes' estimator under improper prior. The experiment is carried out through the following steps:
\textit{
\begin{itemize}
    \item \textbf{Step 1:} For fixed value of $\theta$ generate a sample of size $m$  from UL($\theta$) and store in $\bm{d}$.
    \item \textbf{Step 2:} For $\bm{d}$, calculate $\widehat{\theta}_{ML}$ and thus $(\widehat{\theta}_{ML}-\theta)$, $(\widehat{\theta}_{ML}-\theta)^2$ and stack them into the arrays $\bm{Bias_{ML}}$, $\bm{MSE_{ML}}$, respectively.
    \item \textbf{Step 3:} For $\bm{d}$, calculate $\widehat{\theta}_{I}$ and thus $(\widehat{\theta}_{I}-\theta)$, $(\widehat{\theta}_{I}-\theta)^2$ and stack them into the arrays $\bm{Bias_{I}}$, $\bm{MSE_{I}}$, respectively.
    \item \textbf{Step 4:} Repeat \textbf{Step 1} to \textbf{Step 4} for $N=1000$ times.
    \item \textbf{Step 5:} Take average of the elements for each array: $\bm{Bias_{ML}}$, $\bm{MSE_{ML}}$, $\bm{Bias_{I}}$, $\bm{MSE_{ML}}$ to get bias and MSE of $\widehat{\theta}_{ML}$, bias and MSE of $\widehat{\theta}_{I}$.
\end{itemize}
}
\noindent The results for different combinations of $\theta$ and $m$, are presented in Table 1. From the findings of Table 1, it is quite evident that, $\widehat{\theta}_{ML}$ performs better than $\widehat{\theta}_{I}$.\\

Simulation results reported in Table 3 regarding estimation of $\mathcal{R}$ are also obtained using the above steps incorporating obvious extensions. It is worth observing that unlike $\widehat{\theta}_{I}$, $\widehat{\mathcal{R}}_{I}$ beats $\widehat{\mathcal{R}}_{ML}$. This may be attributed to the complex form of $\widehat{\mathcal{R}}_{ML}$. On the contrary, $\widehat{\mathcal{R}}_{I}$ is less sensitive to the form of $\mathcal{R}$ and hence, should be preferred over MLE in the absence explicit prior knowledge.\\

To assess the performance of Bayes' estimators for $\theta$ under different prior we proceed through the following steps:
\textit{
\begin{itemize}
    \item \textbf{Step 1:} Fix hyper-parameters $\alpha$, $\beta$, $p$.
    \item \textbf{Step 2:} Draw $\theta\sim WG(\alpha,\beta,p)$ using the algorithm in section 2.2.3. 
    \item \textbf{Step 3:} Generate sample of size $m$ from UL($\theta$) and store in $\bm{d}$.
    \item \textbf{Step 4:} For $\bm{d}$, compute $\widehat{\theta}_{A}$ with accurate prior and hence stack $(\widehat{\theta}_{A}-\theta)$, $(\widehat{\theta}_{A}-\theta)^2$ into $\bm{Bias_A}$ and $\bm{MSE_A}$, respectively.
    \item \textbf{Step 5:} For $\bm{d}$, compute $\widehat{\theta}_{M}$ with miss specified prior and hence stack $(\widehat{\theta}_{M}-\theta)$, $(\widehat{\theta}_{M}-\theta)^2$ into $\bm{Bias_M}$ and $\bm{MSE_M}$, respectively.
    \item \textbf{Step 6:} For $\bm{d}$, compute $\widehat{\theta}_{I}$ with flat prior and hence stack $(\widehat{\theta}_{I}-\theta)$, $(\widehat{\theta}_{I}-\theta)^2$ into $\bm{Bias_I}$ and $\bm{MSE_I}$, respectively.
    \item \textbf{Step 7:} Repeat \textbf{Step 2} to \textbf{Step 6} $N=1000$ times.
    \item \textbf{Step 8:} Take average of the elements for each array: $\bm{Bias_{A}}$, $\bm{MSE_{A}}$, $\bm{Bias_{M}}$, $\bm{MSE_{M}}$, $\bm{Bias_{I}}$, $\bm{MSE_{I}}$  to get bias and MSE of $\widehat{\theta}_{A}$, bias and MSE of $\widehat{\theta}_{M}$ and bias and MSE of $\widehat{\theta}_{I}$.
\end{itemize}
}
\noindent As expected $\widehat{\theta}_{A}$ performs the best whereas $\widehat{\theta}_{M}$ beats $\widehat{\theta}_{I}$ in most cases.\\

Simulation results reported in Table 4 regarding estimation of $\mathcal{R}$ are also obtained using the above steps incorporating obvious extensions. It is worth observing that unlike $\widehat{\theta}_{I}$, $\widehat{\mathcal{R}}_{I}$ beats $\widehat{\mathcal{R}}_{M}$. This is due to the joint effect of miss specification for both the parameters. Needless to mention, $\widehat{\mathcal{R}}_{A}$ performs best.
\section{Data Analysis}\label{data}
As mentioned in section 1, here our main interest lies in applying the proposed metric to a real life dataset. For this purpose, we consider basic and safely managed drinking water services data by country available in WHO data repository. Nature of our study is comparative and for this purpose we select two continents viz., North America and South America. Percentage of rural population using at least basic water services (for the year 2015) for $23$ countries in North America and For $12$ in South America are taken and suitable scale transformation (dividing the available percentages by 101) is made to realize them on $(0,1)$, thus avoiding the boundary problem in ML estimation. For North America, data on $9$ countries are not available. With reference to the discussion on section 1, we take South America as region $A$ and North America as region $B$ since number of observations for the latter one is more. Hence the working dataset is:
\begin{eqnarray*}
\bm{x}&=&(0.9505, 0.9901, 0.8911, 0.8515, 0.8218, 0.8812, 0.3960, 0.8317, 0.8713,\\&& 0.9307, 0.6040, 0.8614, 0.9703, 0.9604)
\end{eqnarray*}
\begin{eqnarray*}
\bm{y}&=&(0.9901, 0.7822, 0.8614, 0.9901, 0.8515, 0.7921, 0.9208, 0.9703, 0.7123,\\&& 0.8713, 0.9307, 0.8515)
\end{eqnarray*}
We perform single sameple Kolmogorov-Smirnov goodness of fit test on both $\bm{x}$ and $\bm{y}$ individually to check model adequacy. MLE for $\bm{x}$ and $\bm{y}$ with corresponding K-S test statistic and p-value are reported in Table 5 which validates implementation of the method.\\

In this study, maximum likelihood estimate of $\mathcal{D}(A,B)$ is found to be:
\begin{equation*}
\widehat{\mathcal{D}}(A,B)=-0.25967
\end{equation*}
 with asymptotic $95$\% confidence interval $(-0.57591,0.05657)$. Bayes' estimate of $\mathcal{D}(A,B)$ with improper prior turns out to be $-0.228254$ for chain length $15000$ with burn-in $5000$. Bayes' estimate with informative prior will perform well under reasonable prior specification but we refrain from doing so. Both the estimates reported here are more or less close and enables one to make a comment on the situational disparity of the two continents objectively with probabilistic background.   
\section{Concluding Remarks}\label{conclusion}
Bayes estimation of the unit-Lindley  distribution is thoroughly investigated. Satisfactory results are obtained. The direction of future study may include development of bivariate unit-Lindley distribution to account for dependence which is quite common in many practical applications and related inferential issues should follow. Incorporation of dependence will necessitate a fresh look into the structure of $\mathcal{R}$ and its estimation. As always, new data will come up with new modelling challenges and use of the so called stress-strength reliability in the manner discussed in this article should inspire practitioners to measure socio-economic disparity based on solid support of statistical inference.  

\section*{References}
Abramowitz, M., \& Stegun, I. A. (1965). \textit{Handbook of mathematical functions: with formulas, graphs, and mathematical tables (Vol. 55)}. Courier Corporation.\\
\\
Al-Mutairi, D. K., Ghitany, M. E., \& Kundu, D. (2013). Inferences on stress-strength reliability from Lindley distributions. \textit{Communications in Statistics-Theory and Methods}, 42(8), 1443-1463.\\
\\
Chivers, C. (2015).  https://cran.r-project.org/web/packages/MHadaptive/ MHadaptive.pdf \\
\\
Efron, B. (2012).\textit{ Large-scale inference: empirical Bayes methods for estimation, testing, and prediction (Vol. 1).} Cambridge University Press.\\
\\
Genç,A.I. (2013). Estimation of $P(X>Y)$ with Topp–Leone distribution, \textit{Journal of
Statistical Computation and Simulation}, 83:2, 326-339.\\
\\
Krishna, H., Dube, M., \& Garg, R. (2017). Estimation of $P(Y< X)$ for progressively first-failure-censored generalized inverted exponential distribution.\textit{Journal of Statistical Computation and Simulation}, 87(11), 2274-2289.\\
\\
Maiti, S. S., \& Mukherjee, I. (2018). On estimation of the PDF and CDF of the Lindley distribution. \textit{Communications in Statistics-Simulation and Computation}, 47(5), 1370-1381.\\
\\
Mazucheli, J., Menezes, A. F. B., \& Chakraborty, S. (2019). On the one parameter unit-Lindley distribution and its associated regression model for proportion data. \textit{Journal of Applied Statistics}, 46(4), 700-714.\\
\\
Nadar, M., Kızılaslan, F., \& Papadopoulos, A. (2014). Classical and Bayesian estimation of $P (Y<X)$ for Kumaraswamy's distribution. \textit{Journal of Statistical Computation and Simulation}, 84(7), 1505-1529.\\
\\
Robert, C. (2007). \textit{The Bayesian choice: from decision-theoretic foundations to computational implementation.} Springer Science \& Business Media.\\
\\
Zakerzadeh, H., \& Dolati, A. (2009). Generalized Lindley Distribution. \textit{Journal of Mathematical Extension}.

\section*{Appendix: Tables}

\begin{center}
\begin{table}[h!]
\centering
\vspace{1cm}
	\caption{Bias (mean-squared error) of different estimators for $\theta$.}
	\label{tab:simul}
	\begin{tabular}{cccc}
		\hline
		$m$ & $\theta$ & ${\widehat{\theta}_{ML}}$ & ${\widehat{\theta}_{I}}$ \\ \hline
		\multirow{5}{*}{20} & 9.00 &0.445(4.377)  &0.872(5.374)   \\
		& 2.33 &0.097(0.225)  &0.186(0.271)  \\
		& 1.50 &0.049(0.083)  &0.101(0.097)   \\
		& 4.00 &0.166(0.743)  &0.336(0.901)   \\
		& 0.43 &0.011(0.006)  &0.023(0.006)  \\ \hline
		\multirow{5}{*}{40} & 9.00 &0.204(1.915)  &0.411(2.136)    \\
		& 2.33 &0.041(0.098)  &0.085(0.108) \\
		& 1.50 &0.023(0.038)  &0.052(0.041)   \\
		& 4.00 &0.080(0.335)  &0.163(0.372)   \\
		& 0.43 &0.006(0.003)  &0.012(0.003)  \\ \hline
		\multirow{5}{*}{60} & 9.00 &0.153(1.230) &0.290(1.330)   \\
		& 2.33 &0.033(0.067)  &0.062(0.071)  \\
		& 1.50 &0.014(0.023)  &0.031(0.025)   \\
		& 4.00 &0.062(0.209)  &0.116(0.226)   \\
		& 0.43 &0.005(0.002)  &0.008(0.002)  \\ \hline
		 
	\end{tabular}%
\end{table}
\end{center}

\begin{table}[]
\centering
\vspace{1cm}
	\caption{Bias (mean-squared error) (in order of $10^{-4}$) of Bayes' estimators for $\protect\theta$ with different priors.}
	\label{tab:simul}
	\begin{tabular}{cccccc}
		\hline
		$m$ & $(\alpha,\beta,p)$&
		$(\alpha^\prime,\beta^\prime,p^\prime)$& ${\widehat{\theta}_{A}} $& ${\widehat{\theta}_{M}}$&${\widehat{\theta}_{I}}$ \\ \hline
\multirow{3}{*}{20} & (1,1,1)&(3.1,1.9,2) & 331.324(345.607) &-154.985(367.185)  &795.456(718.155)  \\
		& (2.3,4.7,3) &(1.2,3.6,4) &251.977(137.109)   &684.887(228.899)   &553.082(227.156)  \\
		& (6,5.2,4)&(6,4,5)  &116.380(58.499) &69.801(58.227) &456.371(103.095) \\\hline
		
		\multirow{3}{*}{40} & (1,1,1)&(3.1,1.9,2) &185.712(135.623)  &-87.022(187.999)  &384.747(276.104)  \\
		& (2.3,4.7,3) &(1.2,3.6,4) &119.179(70.325)   &337.114(84.453)   &375.501(126.903)  \\
		& (6,5.2,4)&(6,4,5)  &55.452(30.878) &24.091(31.667) &214.942(41.407) \\\hline
		
		\multirow{3}{*}{60} & (1,1,1)&(3.1,1.9,2) &89.539(98.716)  &-121.197(116.965)  &193.227(118.203)  \\
		& (2.3,4.7,3) &(1.2,3.6,4) &65.139(52.178)   &181.075(56.888)   &211.592(61.833)  \\
		& (6,5.2,4)&(6,4,5)  &82.809(19.451) &13.223(19.755) &165.456(28.34) \\\hline

	\end{tabular}%
\end{table}

\begin{table}[]
\centering
\vspace{1cm}
	\caption{Bias (mean-squared error) (in order of $10^{-4}$) of different estimators for $\mathcal{R}$.}
	\label{tab:simul}
	\begin{tabular}{cccccc}
		\hline
		$(m,n)$ & $(\theta_1,\theta_2)$&$\mathcal{R}$& ${\widehat{\mathcal{R}}_{ML}}$ &${\widehat{\mathcal{R}}_{I}}$& CI \\ \hline
\multirow{5}{*}{(20,20)} &(9.00,4.00)&0.285 &55.731(45.359)  &92.353(44.959)  &(0.192,0.389)  \\
       &(2.33,4.00) &0.659 &-18.103(47.751)  &-47.524(46.538)  &(0.448,0.867)  \\
		&(1.50,1.00) &0.366 &-21.088(56.039) &4.405(54.099)  &(0.240,0.489)  \\
		 &(4.00,0.67) &0.084 &15.438(7.490)  &48.785(8.000)  &(0.049,0.122)  \\
		&(0.43,1.00) &0.771 &-66.636(36.076) &-110.326(36.224) &(0.545,0.983) \\\hline
		
		\multirow{5}{*}{(40,20)} &(9.00,4.00)&0.285&51.471(33.882)  &102.787(34.420)  &(0.215,0.365)   \\
       & (2.33,4.00) &0.659 &15.123(38.630)  &16.360(37.597)  &(0.462,0.859)  \\
		&(1.50,1.00) &0.366  &33.740(43.125) &76.615(42.720)  &(0.268,0.472)  \\
		 &(4.00,0.67) &0.084  & 41.476(6.174) &76.189(6.881)  &(0061,0.115)\\
		&(0.43,1.00) &0.771  &2.305(27.529) &-13.479(26.966) &(0.558,0.983) \\\hline

		\multirow{5}{*}{(40,40)} & (9.00,4.00)&0.285& 3.809(21.539)  &23.897(21.362)   &(0.216,0.354)   \\
       & (2.33,4.00) &0.659  & 11.378(26.327) &-4.726(25.879)  &(0.512,0.809)  \\
		& (1.50,1.00) &0.366  &50.071(31.136) &63.263(30.793)  &(0.282,0.461)  \\
		 & (4.00,0.67) &0.084 & 20.052(3.769) &37.273(3.963)  &(0.060..0112)  \\
		& (0.43,1.00) &0.771 &-24.496(18.560) &-47.450(18.608) &(0.612,0.924) \\\hline

	\end{tabular}%
\end{table}
\begin{landscape}
\begin{table}[]
\centering
\vspace{1cm}
	\caption{Bias (mean-squared error) (in order of $10^{-5}$) of Bayes' estimators for $\mathcal{R}$ with different priors.}
	\label{tab:simul}
	\begin{tabular}{cccccccc}
		\hline
		$(m,n)$ & $(\alpha_1,\beta_1,p_1)$ &$(\alpha_2,\beta_2,p_2)$ & $(\alpha^\prime_1,\beta^\prime_1,p^\prime_1)$&$(\alpha^\prime_2,\beta^\prime_2,p^\prime_2)$ & ${\widehat{\mathcal{R}}_{A}}$ &${\widehat{\mathcal{R}}_{M}}$&${\widehat{\mathcal{R}}_{I}}$ \\ \hline
\multirow{3}{*}{(20,20)} & (1,1,1)&(1,1,1)   &(3.1,1.9,2)&(2.2,3.2,2) &95.127(268.202)  &62.538(230.162)  &33.884(269.368)  \\
		& (2.3,4.7,3)&(5.2,3.6,4) &(1.2,3.6,4)&(7.3,5.0,3)  &-371.875(426.978)   &-3086.691(519.702)   & 92.753(417.404)    \\
		& (6,5.2,4)&(6,4,5) &(5,4.4,2)&(7.1,5.3,1)  &-257.403(432.304)  &-2303.736(467.789) &-119.287(446.023)  \\\hline

\multirow{3}{*}{(40,20)} & (1,1,1)&(1,1,1) &(3.1,1.9,2)&(2.2,3.2,2)&35.341(165.780)   &51.560(156.743)  &-50.735(148.472)  \\
		& (2.3,4.7,3)&(5.2,3.6,4) &(1.2,3.6,4) &(7.3,5.0,3) &-33.006(211.139)   &-1765.020(250.514)   & -44.443(217.053)    \\
		& (6,5.2,4) &(6,4,5)&(5,4.4,2)&(7.1,5.3,1)  &104.657(221.095)  &-1479.458(240.938)  &-144.490(236.988)   \\\hline
		
\multirow{3}{*}{(60,60)} & (1,1,1)&(1,1,1) &(3.1,1.9,2)&(2.2,3.2,2)&31.100(96.994)   &-15.983(92.742)  &-117.199(98.549)  \\
		& (2.3,4.7,3)&(5.2,3.6,4) &(1.2,3.6,4)&(7.3,5.0,3) &13.014(154.847)   &-1246.592(183.749)   & 69.951(156.650)    \\
		& (6,5.2,4) &(6,4,5)&(5,4.4,2)&(7.1,5.3,1)  &-258.978(138.277)  &-836.536(159.317)  & -185.032(146.759)  \\\hline
	\end{tabular}%
\end{table}
\end{landscape}

\begin{center}
\begin{table}[h!]
\centering
\vspace{1cm}
	\caption{MLE, K-S Statistic and p-value for real data}
	\label{tab:simul}
	\begin{tabular}{cccc}
		\hline
		Data & MLE& K-S statistic & p-value\\
		\hline
		$\bm{x}$ & 0.11282 & 0.25394 &0.2772\\
		
		$\bm{y}$ & 0.07916 & 0.17452 &0.8582\\
		\hline
	\end{tabular}%
\end{table}
\end{center}

(A preprint of the earlier version of this original research article is available at: \textit{https://arxiv.org/abs/1904.06181})
\end{document}